\documentclass[aps,prl,twocolumn,groupedaddress]{revtex4-1}
\usepackage{url} 
\usepackage{lineno}
\usepackage{amssymb}
\usepackage{amsmath}
\usepackage{soul}
\usepackage{natbib}
\usepackage{graphicx}
\usepackage{array,booktabs}
\newcolumntype{L}{@{}>{\kern\tabcolsep}l<{\kern\tabcolsep}}
\usepackage{colortbl}
\usepackage{xcolor}
\begin{document}

\title{THMC instabilities and the Multiphysics of Earthquakes } 

\author{K. Regenauer-Lieb}
\affiliation{School of Minerals and Energy Resources Engineering, UNSW, Sydney, NSW 2052 Australia }
\author{ M. Hu}
\affiliation{Department of Civil Engineering, The University of Hong Kong, Hong Kong }
\affiliation{School of Minerals and Energy Resources Engineering, UNSW, Sydney, NSW 2052 Australia }
\author{C. Schrank}
\affiliation{Science and Engineering Faculty, Queensland University of Technology,  Brisbane, QLD, 4001, Australia }

\begin{abstract}
Theoretical approaches to earthquake instabilities propose shear dominated instabilities as a source mechanism. Here we take a fresh look at the role of possible volumetric instabilities preceding a shear instability. We investigate the phenomena that may prepare earthquake instabilities using coupling of Thermo-Hydro-Mechano-Chemical (THMC) reaction-diffusion equations in a THMC diffusion matrix.  We show that the off-diagonal cross-diffusivities can give rise to cross-diffusion slow (no inertia) pressure waves and propose that a resonance phenomenon of the long-wavelength limit causes constructive interference and triggers the seismic moment release.
\end{abstract}

\maketitle

\section{Introduction}
We present the hypothesis that the universality of cross-diffusion phenomena based on multiscale, multiphysics coupling underpins the scale invariance of the Gutenberg, Richter frequency-magnitude relationship of earthquakes \citep{Sethna}. We propose a new mathematical approach for coupling the rates of chemical reactions, failure and fluid flow instabilities and thermo-mechanical instabilities of materials. The approach gives theoretical insights into the processes that are commonly described by empirical relationships. 

In our multiphysics approach to earthquakes we consider local chemical reactions and mass exchange processes as the smallest thermodynamic driving force triggering a thermodynamic flux. The approach emphasises the important role of emergent patterns (microstructure) triggering energy-geometry interactions that cascade through space and time via the link of THMC cross-diffusion coefficients introduced in this paper. These coefficients can be activated in far from equilibrium conditions, which allows coupling between the scales. The important factor of cross-scale coupling by cross-diffusion was overlooked in earlier reviews of multiscale coupling and multiphysics approaches \citep{JCSMD1, JCSMD2} to earth physics problems. We propose here that earthquakes can appear if the large-scale geodynamic driver fires the full cascade of THMC cross-diffusion coefficients through reaching a critical magnitude of reactive THMC source terms. 

\section{ Multiscale cross-diffusion model}

The conventional  definition of thermodynamic forces and fluxes in a THMC-system is given in Table 1. Thermodynamic forces are understood as the spatial gradients of a THMC-system. They drive time gradient thermodynamic fluxes. At the molecular scale, the thermodynamic force is, for instance, the gradient of a chemical species, which drives a chemical flux (C) parameterized by Fick’s law, where ${{D}_{C}}$ is the chemical diffusivity. Such chemical reactions can in turn become a driver for the next scale up in the multiscale system. For example, a chemical fluid release reaction in an HMC-coupled system can create mechanical pressure gradients due to the associated volume change and fluid release/precipitation, which in turn induces mechanical (M) stress diffusion ${{D}_{M}}$ with a Beltrami-Michell-type compatibility condition \citep{stressdiffusion} and Darcy-type fluid (H)  diffusion ${{D}_{H}}$. In the mechanical sense, the chemical reaction creates an overpressure ${{\bar{p}}_{M}}$ in the matrix \citep{Hu2019b}. The endo-/exo-thermal chemical reaction also induces a gradient in temperature which in turn drives a thermal Fourier flux (T), with thermal diffusivity ${{D}_{T}}$.

\begin{table}
\centering
\caption{THMC Thermodynamic Forces and Fluxes and Conservation Laws ($\mathcal{D}(\cdot)/\mathcal{D}t$ denotes the material derivative considering the convective effects) }
\includegraphics[width=.5\textwidth]{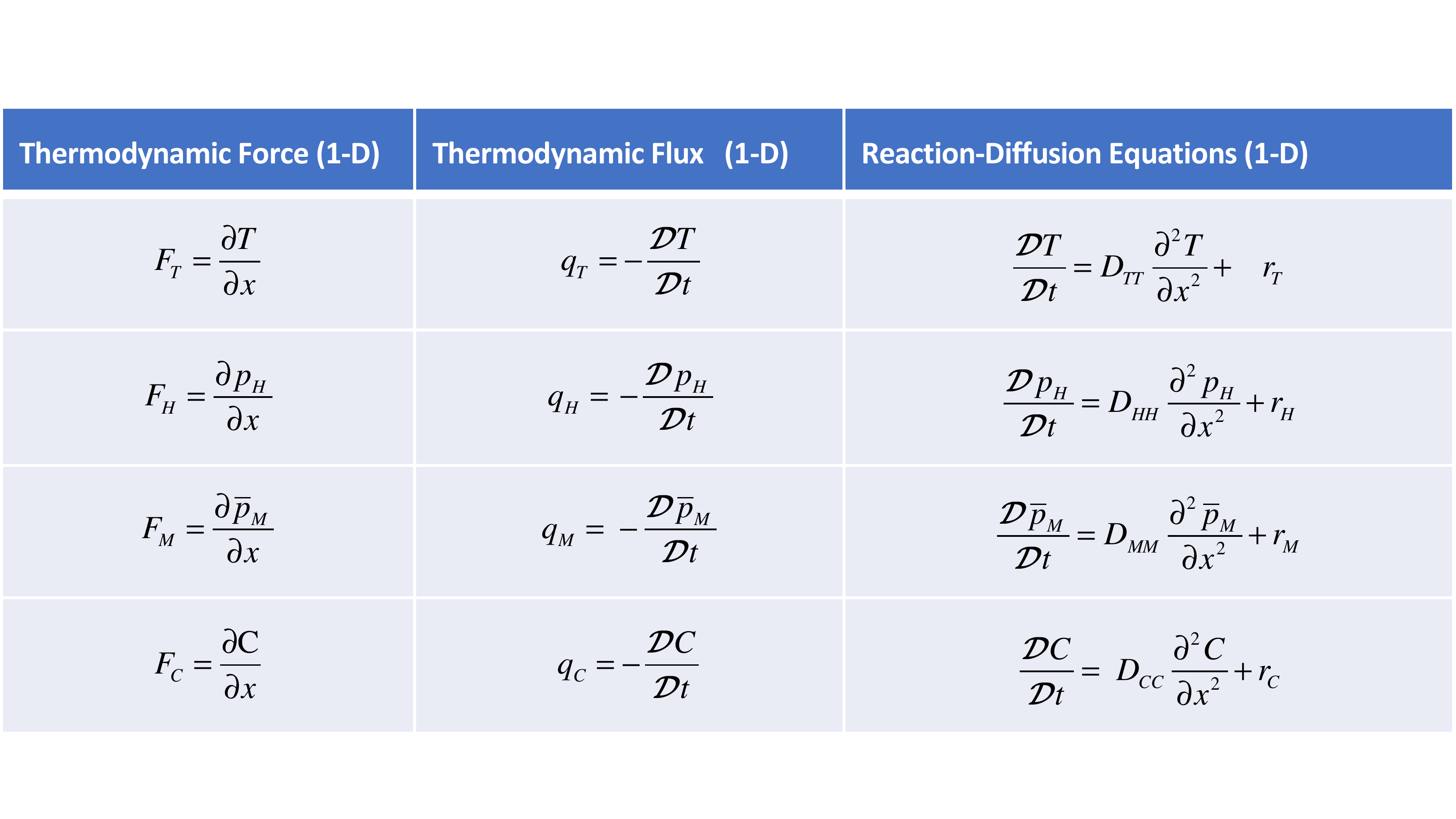}
\label{tab:table1}
\end{table}

The afore mentioned cross-scale coupling introduces source/sink terms in the individual conservation laws identified by ${{r}_{T}},{{r}_{H}},{{r}_{M}}$ and ${{r}_{C}}$ respectively. The conservation laws must also be extended to allow a multitude of THMC processes to occur simultaneously, which can introduce cross-diffusion fluxes. 

\subsection{THMC cross-diffusion model formulation }
In a chemical system with two species, for instance, cross-diffusion is the phenomenon, in which a flux of species $A$ is induced by a gradient of species $B$ \citep{Vanag}. In more general THMC terms, cross-diffusion is the phenomenon where a gradient of one generalised thermodynamic force of species ${C}_{j}$ drives another generalised thermodynamic flux of species ${C}_{i}$, described by \citep{Manning1970}
\begin{equation}
\label{eq:crossdiff2}
\frac{\mathcal{D}{{C}_{i}}}{\mathcal{D}t}={{D}_{ii}}{{\nabla }^{2}}{{C}_{i}}+\sum\limits_{j\ne i}{{{D}_{ij}}}{{\nabla }^{2}}{{C}_{j}}+{{r}_{i}} \quad .
\end{equation}

The species  $j$ is identified as the cross diffusion species other than the species $i$ . Introducing a fully populated ($N \times N$) diffusion matrix ${{\mathbf{D}}_{ij}}$, equation (\ref{eq:crossdiff2}) can also be written as

\begin{equation}
\label{eq:crossdiffusionmatrix}
\frac{\mathcal{D}{{C}_{i}}}{\mathcal{D}t}=\sum\limits_{k=1}^N {{{D}_{ij}}}{{\nabla }^{2}}{{C}_{k}}+{{r}_{i}} \quad ,
\end{equation}
The diffusion matrix ${{\mathbf{D}}_{ij}}$  for the equivalent THMC processes is

\begin{equation}
{{\mathbf{D}}_{ij}}= 
\begin{bmatrix}  
   {\textcolor{red}{{D}_{TT}}} & {{L}_{TH}} & \textcolor{blue}{{{L}_{TM}}} & {{L}_{TC}}  \\
   {{L}_{HT}} & \textcolor{red}{{{D}_{HH}}} & {{L}_{HM}} & {{L}_{HC}}  \\
   \textcolor{blue}{{{L}_{MT}}} & {{L}_{MH}} & \textcolor{red}{{{D}_{MM}}} & {{L}_{MC}}  \\
   {{L}_{CT}} & {{L}_{CH}} & {{L}_{CM}} & \textcolor{red}{{{D}_{CC}}}  \\
\end{bmatrix} .
\end{equation}

The classical (self-)diffusive length scale of each THMC process is defined by $\sqrt{D_{ii}t}$. The reactive source source term $r_i$, corresponding to each thermodynamic flux ${C}_{i}$, can be zero or non-zero. The approach presented here introduces a new cross-scale coupling length scale that provides a link between the self-diffusion length scales of the THMC processes. Although not essential to the formulation we propose that this cross-diffusive length scale $\sqrt{D_{ij}t}$ is quantized and defined by discrete material properties which can be evaluated by measuring the velocity of cross-diffusion waves \citep{Regenauer-Lieb2019b}. These (quantized) cross-diffusion coefficients link the gradient of a thermodynamic force ${C}_{j}$ of one THMC process to the flux of another kind.
 
Under such circumstances, the criterion for nucleation of diffusion waves is fully characterized by the diffusion matrix ${{\mathbf{D}}_{ij}}$ with the diagonal elements ${{D}_{ii}}$ describing the normal self-diffusion and the off-diagonal elements the cross-diffusion processes enabling coupling across scales. For consistency with the second law, all eigenvalues of the diffusion matrix must be real and positive, and hence the determinant of the diffusion matrix $\det \left( \mathbf{D}_{ij} \right)>0$ as well as the trace of the diffusion matrix must be larger than zero.

The criterion for instability can now be expressed by two matrices, the self- and cross-diffusion matrices, respectively. The cross-diffusion matrix must satisfy the Onsager-Casimir symmetry condition which acknowledges antisymmetry ${{D}_{ij}}=\pm{{D}_{ji}}$ \citep{Onsager,Casimir}. Onsager \citep{Onsager} assumed that the parameters characterizing the system have identical sign in their kinetic coefficients while Casimir \citep{Casimir} extended Onsager’s considerations to include kinetic parameters that have opposite sign as can be the case for multiphysics cross-diffusion coupling.  Any local perturbation that leads to complex eigenvalues of ${\mathbf{D}_{ij}}$ must lead to oscillatory instabilities relaxing to the equilibrium state \citep{Vanag}. This statement is at the heart of the nucleation of the cross-diffusional wave phenomenon. We will show in the following sections that the above described cross-diffusion formulation \citep{Vanag} can be extended to develop a generic multiphysics and multiscale THMC coupling approach to earthquake instabilities.  This formulation implies a coupled cascade of THMC feedbacks over multiple diffusional length scales honouring the reciprocal multiscale interplay of thermodynamic forces and fluxes. 

\subsection{Cross-diffusion and its role in coupling of instabilities across scales}
By extending the diagonal diffusion matrix through the cross-diffusion coefficients in equation~(\ref{eq:crossdiffusionmatrix}), a new class of mechanical instabilities is revealed that incorporates chemical instabilities at its lowest scale \citep{Regenauer-Lieb2019b}.  For the simple case of hydro-mechanical (HM) coupling  we have recently reported discovery of the cross-diffusional pressure wave phenomenon \citep{Hu2019b}. 

The important element of cross-diffusion waves for earthquake physics is their capability to link one thermodynamic force with a thermodynamic flux at a different scale, thus synchronising the dynamics over vastly different diffusional time and length scales. This important aspect of earthquake physics was previously overlooked. The approach raises the possibility that dissipative waves can be detected prior to earthquake instabilities. 

\section{Application to earthquake physics}
Theoretical approaches to the physics of earthquake instabilities originally were conceived as a shear instability on a frictional sliding surface \citep{Brace1966}. The role of pressure on the dynamics of the slider was derived empirically by  laboratory experiments defining a rate-and-state variable friction law \citep{Dieterich72,Dieterich1981,Ruina83,Tse1986, rice2001rate}. Instabilities through shear heating feedback were later considered to play an important role \citep{Ogawa, KRL98, Braeck}. Additionally, a thermally induced fluid pressurization term was found to be an important component for accelerated creep \citep{Vardoulakis2001,Rice2006}. Another important ingredient of the earthquake instability was thought to be the coupling of scales, where at least two different processes, operating at different time and length scales, interact \citep{Ohnaka}. The approach presented here summarizes these effects in the matrix ${{\mathbf{D}}_{ij}}$ and enables upscaling via renormalization group theory \citep{JCSMD1, Hanasoge}.

\subsection{Cross-diffusion waves}
A fully populated diffusion matrix provides the opportunity to extend  the postulate of coupling different scale processes in earthquake mechanics \citep{Ohnaka}. We present  a high-level application of the basic principles in order to transfer knowledge from other disciplines to this new field of research (e.g. mathematical biology, computational chemistry, quantum optics).  We use closed form solutions developed in these fields, supplemented by our own solutions  \citep{JCSMD2, Veveakis2015cnoidal}, for schematic illustrations and in-depth discussions of how the cross-diffusion concept may be applied to the earthquake phenomenon. 

Exact analytical propagating wave solutions of the cross-diffusion wave phenomenon are rare. The FitzHugh-Nagumo oscillator \citep{Zemskov} is by far the best analysed prototype of an excitable system with short lived-spikes and a slow recovery. The FitzHugh-Nagumo cross-diffusion model has originally been developed to describe the propagation of electrical impulses in nerves. We use  a generic application of this oscillator to illustrate the potential feedback of just two THMC processes and the conditions that prepare the formation of a future fault plane. The real waveforms of the transient cross-diffusion waves are expected to be more complex.  

\subsubsection{Cross-diffusion wavenumbers}
The wavenumbers of the cross-diffusion waves are controlled by the competition of the diffusion processes determining the diffusion length and number of possible oscillations per unit length. Of particular interest for the earthquake instabilities is the resonant mode with the longest possible wavelength. 

Below, we discuss the results of a 1-D linear stability analysis for just two coupled reaction-diffusion equations to illustrate the role of the wavelength (or its inverse the wavenumber) for the nucleation of cross-diffusion waves. Laboratory evidence of cross-diffusion waves with HM coupling and a detailed derivation of the equations can be found in ref. \cite{Hu2019b}. For HM coupling, we set thermal and chemical self- and cross-diffusion coefficients to zero in equation (\ref{eq:crossdiffusionmatrix}) and using a single source term characterized by the ratio of the self-diffusivities of the two processes:

\begin{subequations}
\label{eq:crossdiff1}
\begin{align}
\frac{\mathcal{D}{{C}_{1}}}{\mathcal{D} t}=D_{HH} \frac{\partial ^2 C_1}{\partial x^2}+D_{HM} \frac{\partial ^2 C_2}{\partial x^2} \\[0.2in]
\frac{\mathcal{D}{{C}_{2}}}{\mathcal{D} t}=D_{MM} \frac{\partial ^2 C_2}{\partial x^2}+D_{MH} \frac{\partial ^2 C_1}{\partial x^2}-{\lambda C_2} 
\end{align}
\end{subequations}

where $\lambda={D_{MM}}/{D_{HH}}$, is a non-dimensional parameter denoting the ratio of mechanical diffusivity over the fluid diffusivity.
The response of the system to wave perturbations is tested by applying a small plane wave perturbation of the type

\begin{equation}
\label{eq:LSA2}
\begin{bmatrix} C_1  \\[0.3em] C_2 \end{bmatrix} =
\begin{bmatrix} C_1 \exp \left( i k x+\omega t \right) \\[0.5em]
 C_2  \exp \left( i k x+\omega t \right) \end{bmatrix}  ,
\end{equation} 

where $k$ is the wavenumber and $\omega$ the angular frequency. The perturbation yields condition for instability when \citep{Hu2019b} 

\begin{equation}
\label{eq:LSA3}
  (D_{HH} - D_{MM} -\frac{D_{MM}}{D_{HH}~k^2} )^2  + 4 {{D}_{HM}} {{D}_{MH}} < 0  .
\end{equation}

A cross-diffusional wave can nucleate if ${{D}_{HM}}$ and ${{D}_{MH}}$  are nonzero, of opposite sign and sufficiently large to overcome the self-diffusion processes in the first term, which is always positive. The instability criterion also suggests a critical lower limit  $k_{min}$ (longest wavelength) for which no cross-diffusion wave can nucleate as the first term becomes dominant. This minimum wavenumber (longest wavelength) can be constrained further by the long wavelength limit of the system size. We argue that this size is a fraction of the system size, defined by long-wavelength  Korteweg-deVries equation solution \citep{JCSMD2, Veveakis2015cnoidal} where the cross-diffusion coefficients are adiabatically eliminated. This long-wave soliton solution allows constructive interference of the cross-diffusional waves in normal resonance mode. 

\subsubsection{Resonance on the lowest possible wavenumber}
\begin{figure}
\centering
\includegraphics[width=.5\textwidth]{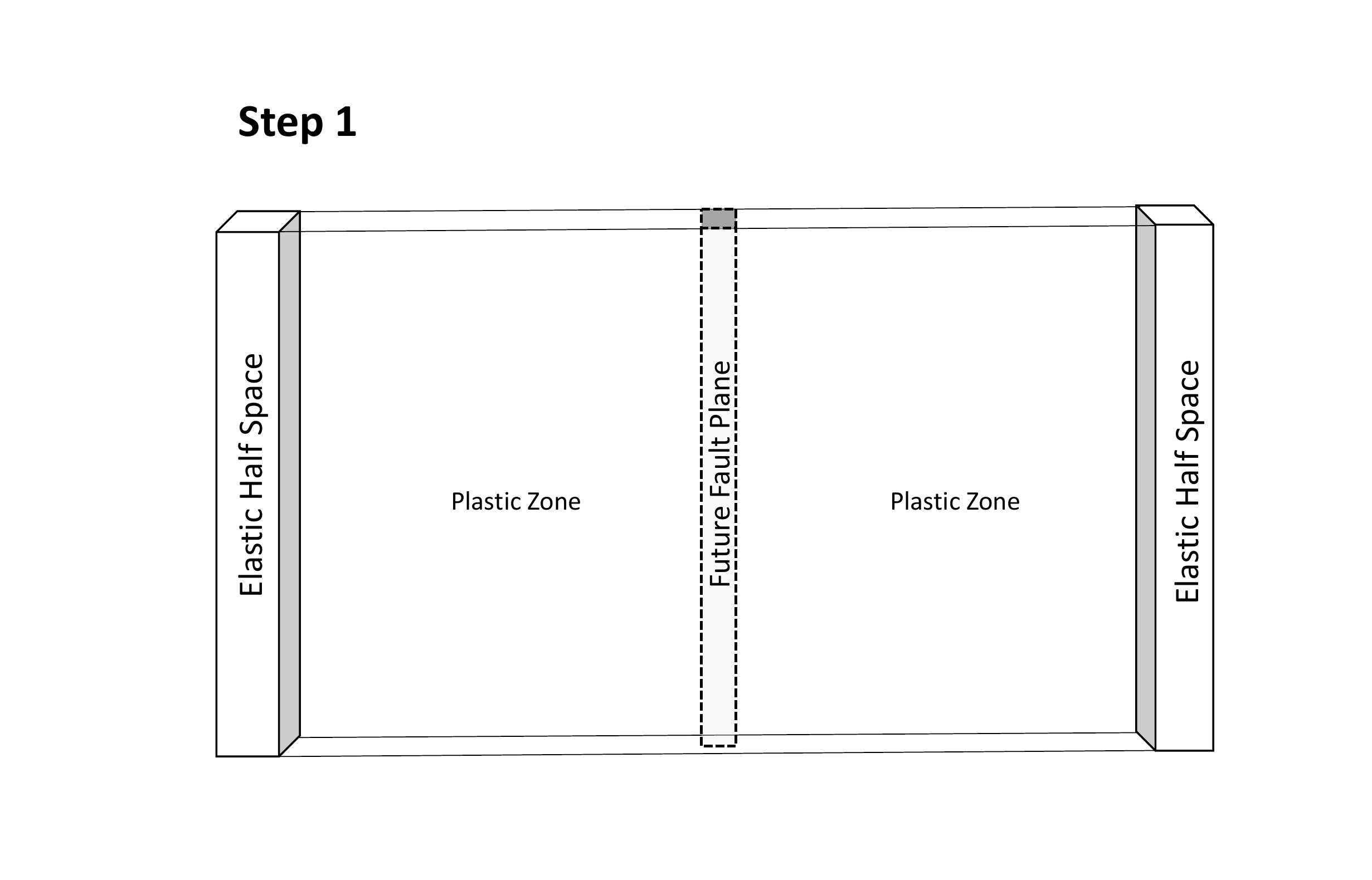}
\caption{Step 1 development of a plastic zone. The size of the zone is dependent on the boundary conditions and the micro-physics accommodating plastic deformation. We assume in our discussion that the zone is deforming pervasively by a power-law $r_2={- \frac{D_M}{D_H} C_2}^m$ . }
\label{fig:Step1}
\end{figure}

\begin{figure}
\centering
\includegraphics[width=.5\textwidth]{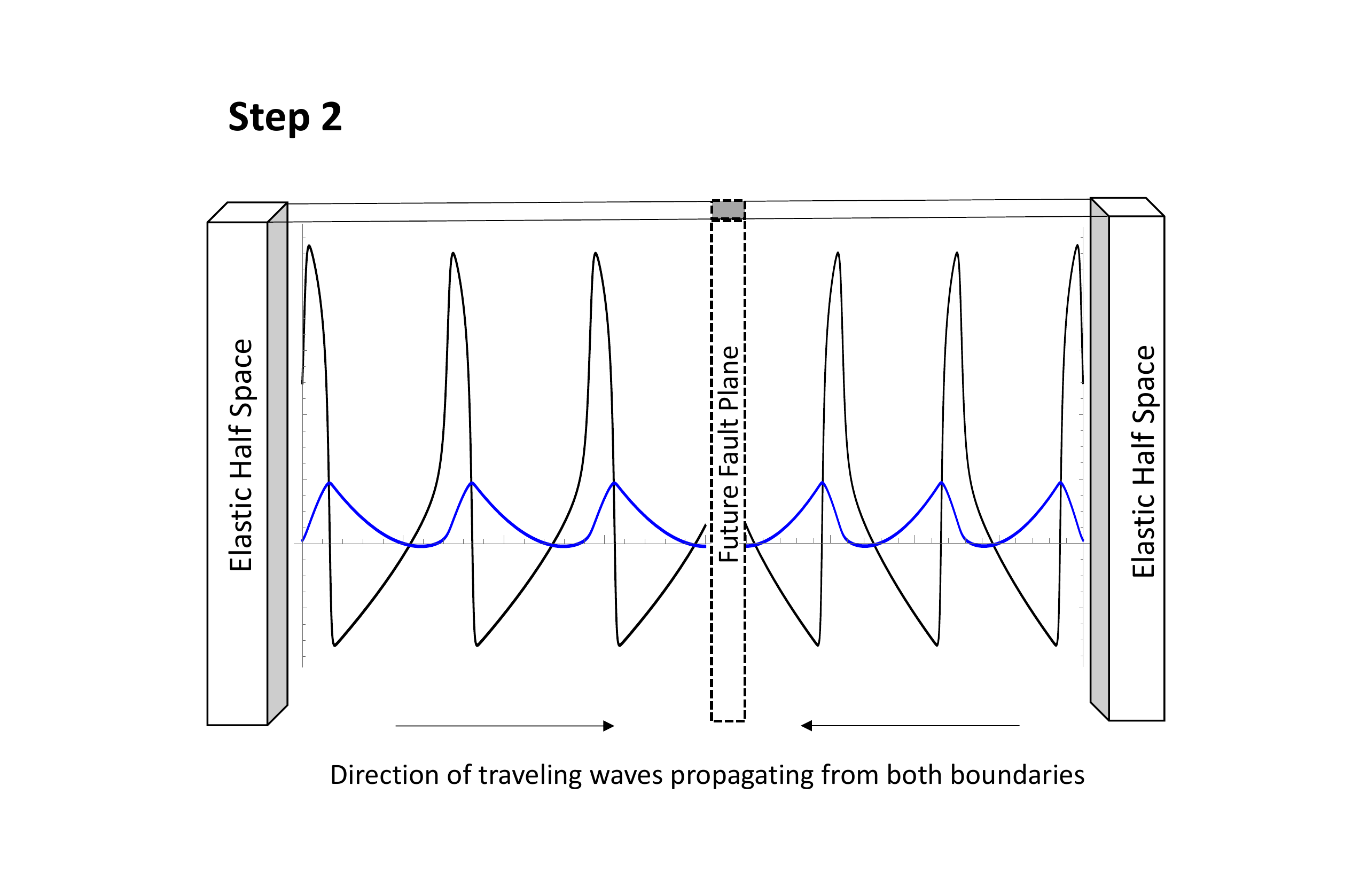}
\caption{For critical conditions specified in equation (\ref{eq:LSA3}) excitable self-oscillatory cross-diffusion waves propagate from opposite sides towards the future fault plane. Here, we use the analytical solution of the piecewise linear approximation of the FitzHugh-Nagumo cross-diffusion wave  \citep{Zemskov} for illustration. For the earthquake cross-diffusion $C_1$ (black curve) represents the oscillation of fluid pressure  and $C_2$ (blue curve) the pressure oscillation including  the source term of the deforming solid matrix. A snapshot is shown just before collision on the future fault plane. The consequence of collisions of cross-diffusion waves is discussed in the section on waveforms.}
\label{fig:Step2}
\end{figure}

\begin{figure}
\centering
\includegraphics[width=.5\textwidth]{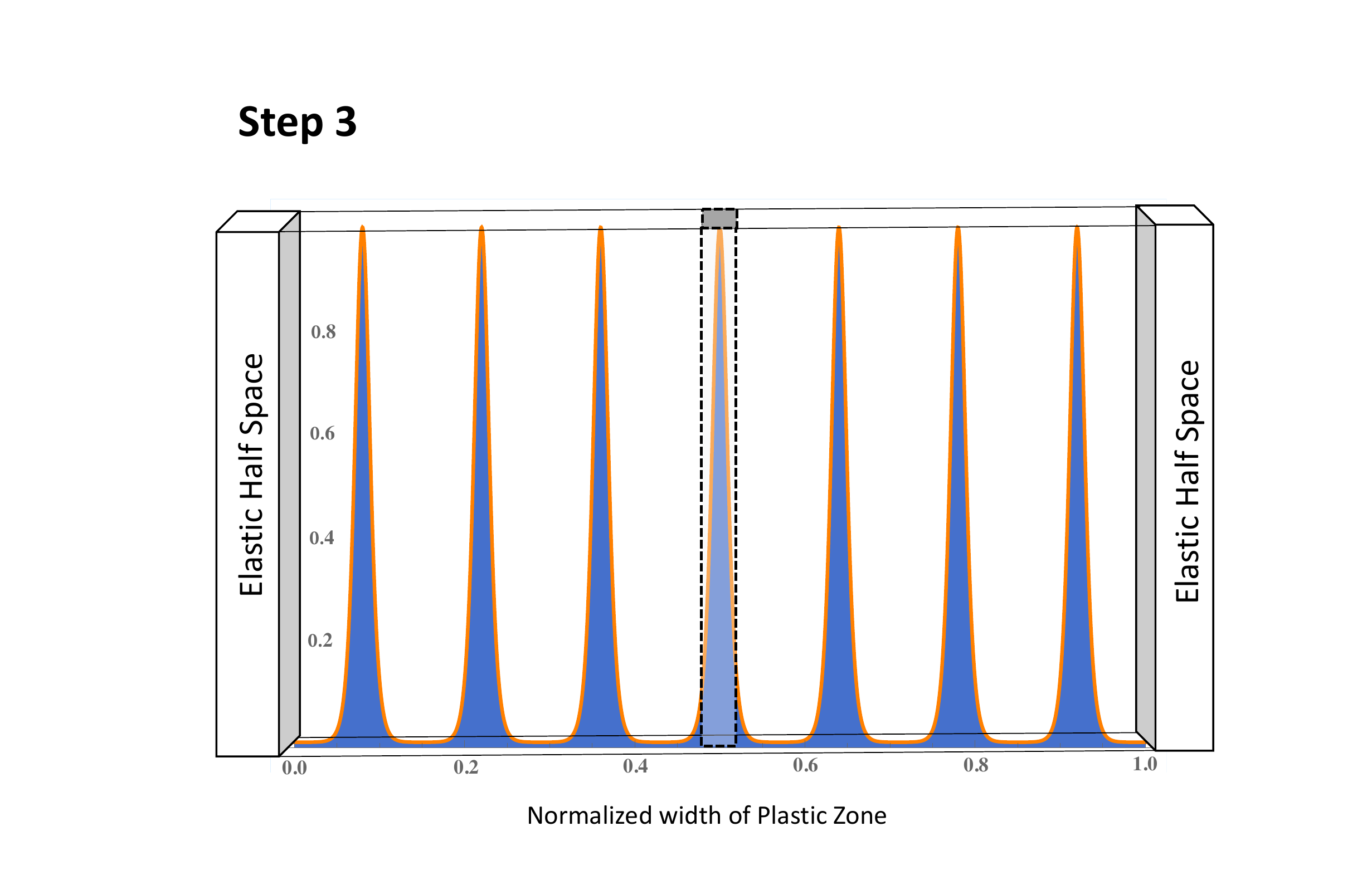}
\caption {In the case that the instability proceeds slowly, no earthquake occurs and no central fault plane can be identified. The lowest wavenumber on the  $k_{min}$ resonance P-wave mode can fully develop and the static solution can be reduced to lower dimension by considering only the self-diffusion coefficients. The solution is then described by the Korteweg-de Vries equation \citep{JCSMD2, Veveakis2015cnoidal} and a strictly periodic pattern of overpressured fluid channels (non-dimensional overpressure labelled on the ordinate) is expected. A good example is the periodic fracture pattern stemming from slow diagenetic reactions in unconventional shale gas reservoirs or coal cleats in coal seams \citep{Alevizos2017}.}
\label{fig:Step3}
\end{figure}

In the context of an earthquake instability, shorter wavelength cross-diffusional waves might lead to a small dispersion limit of a coordinated long-wavelength instability. This resonance mode may arise through incident cross-diffusional waves colliding with reflected cross-diffusional waves  in a finite  width around the future fault core.  On the grounds of the above theoretical considerations we postulate a series of evolutionary steps that may lead to this macro-scale instability. These are:

\begin{itemize}
\item{\textbf{Step 1} (illustrated in Figure \ref{fig:Step1}); Upon ongoing geodynamic loading a zone may form, where the rock is loaded past the elastic limit, and the plastic yield stress of the rock is reached. This defines the system size of reflecting cross-diffusion waves.}

\item{\textbf{Step 2} (illustrated in Figure \ref{fig:Step2}); If the criterion in equation (\ref{eq:LSA3}) is met, cross-diffusional waves are expected to initially bounce back and forth inside the plastically deforming domain as wave packets with dominant finite wavenumbers \citep{Biktashev2016, Tsyganov2014},  described in the  next chapter on waveforms. In the hydromechanically coupled problem (equation \ref{eq:crossdiff1}), these waves are propagating fronts of internal volumetric fluid-solid mass exchange and appear as local fluid overpressure. Figure (\ref{fig:Step2}) shows conditions that may lead to the formation of the future fault plane by collision of cross-diffusion waves approaching each other from opposite sides of the fault plane. Collisions of cross-diffusion waves can lead to accumulation of material damage discussed in the section on waveforms.}

\item{\textbf{Step 3} (illustrated in Figure \ref{fig:Step3}); The long wavelength limit of cross-diffusional waves is defined by a fraction of the width of the plastic zone. A large-scale instability is then understood as the longest wavelength normal (resonance) mode to form a standing wave. Earthquakes may occur prior to this situation when the volumetric damage accumulated in Step 2 lubricates a shear instability. For extremely slow deformation the purely volumetric long-wavelength solution of the standing P-wave can fully develop and is shown in Figure \ref{fig:Step3}. We use the analytical solution of the Korteweg-de Vries equation \citep{JCSMD2, Veveakis2015cnoidal} and  emphasize  that the solutions for  Figures \ref{fig:Step2} and \ref{fig:Step3} are for illustrative purposes only.}

\end{itemize}

The earthquake trigger for positive interference necessary for the appearance of a normal mode standing wave is defined by substituting $k_{min}= \sqrt{\frac{D_{MM}}{D_{HH}}}$, obtained from the standing wave-limit, into the instability criterion (\ref{eq:LSA3}):

\begin{equation}
\label{eq:LSA4}
 ( D_{HH} - D_{MM} -1)^2  + 4 {{D}_{HM}} {{D}_{MH}} < 0  .
\end{equation}

This limit also applies for the case of a nonlinear source term defined by a power law, as it coincides with the standing solitary wave of the Korteweg-de Vries equation \citep{JCSMD2,Veveakis2015cnoidal}. 

\subsubsection{Cross-diffusion waveforms}
\label{ch:waveforms}
Cross-diffusion waves bear similar characteristic to quasi-solitons encountered in optical systems where chromatic dispersion is strong leading to anomalous dispersion patterns that, unlike solitons, come in discrete portions \citep{Paschotta}. 

A classification of the solution of equation~(\ref{eq:crossdiffusionmatrix}) for the discussed case of cross-diffusion has been presented recently \citep{Tsyganov2014} and a broader definition of quasi-soliton wave has been adopted. Quasi-solitons feature complex dispersion relationships, where the wave velocity of individual  waves (phase velocity) have different velocity to their smooth envelope wave groups (group velocity). This leads to complex phenomena  similar to envelope solitons in the nonlinear Schr\"odinger equation shown \citep{Tsyganov2014} to be akin to the linear cross-diffusion equation in equation~(\ref{eq:crossdiffusionmatrix}). They consequently differ from the classical soliton solution in that the shape and speed of such waves in the established regime does not depend on initial and boundary conditions and is fully determined by the material parameters of the medium that they travel through.

Wave packet  solutions emerge when the wave energy is concentrated around a finite wavenumber. When the dispersion is weak, the envelope travels at approximately uniform group velocity. The addition of linear cross-diffusion can create strong non-linear dispersion, where - depending on the coefficients - three different wave types have been classified: (1) fixed-shape propagating waves, (2) envelope waves, (3a) multi-envelope waves, and (3b) intermediate regimes appearing as multi-envelope waves propagating as fixed shape most of the time but undergoing restructuring from time to time  \citep{Tsyganov2014}.

The collisional behaviour of these classes of cross-diffusional waves is complicated. When the cross-diffusion wave hits an interface, or another cross-diffusion wave the amplitude of the quasi-soliton changes and there is a temporary diminution of the amplitude or in extreme cases an annihilation. In most cases they recover gradually their original form. This is another feature that is different to true solitons which do not change on impact. In two-dimensional systems additional complexities arise as they can penetrate or break on collision, reflect into different directions and  lead to complex patterns \citep{Tsyganov2014} imprinted in the damage zone of seismogenic faults.

Of specific interest for the earthquake problem is the aspect of the fate of the accelerations carried by the waves \citep{Regenauer-Lieb2019b} for cases where the wave reduces in amplitude on collision or annihilates after collision. In experiments carried out with collapsing puffed rice particles \citep{Ricecrispies2015} cross-diffusion waves were detected \citep{Hu2019b} which release an accoustic emission when annihilating after collision with the top surface. We propose that for a homogeneous plastic zone as shown in Figure (\ref{fig:Step2}), the first wave-collisions on the symmetry axis of the future fault plane converts the energy loss of the cross-diffusion wave into local material damage that will act as a seed to future wave interactions and systematically grows the fault. For the case of heterogeneous materials the collision of cross-diffusion waves with internal or bounding material surfaces would cause an alternative seed for the nucleation of earthquakes.

\section{Discussion and conclusions}
\label{ch:Discussion}
We have introduced a new approach to THMC instabilities using the simple concept of cross-diffusion. The effect of cross-diffusion can, for critical conditions, lead  to the formation of cross-diffusional waves which, using the Helmholtz decomposition of the equation of motion,  can be decomposed into P- and S- cross-diffusion waves \citep{Regenauer-Lieb2019b}. The analysis suggests that cross-diffusion provides a crucial link to allow cross-scale coupling of the multiphysics processes potentially leading to earthquake instabilities. For critical cross-diffusion coefficients and associated reaction terms,  an energy cascade of acceleration cross-diffusion pressure waves (diffusion P-waves), from small scale thermo-chemical (TC) and chemo-mechanical (CM) dissipative waves to meso-scale hydro-mechanical (HM) dissipative waves, can be triggered by a geodynamic driving force. This in turn may nucleate shear-cross-diffusion waves in the form of a thermo-mechanical (TM) shear-wave instability. 

The multiscale THMC-waves initially are low-energy release volumetric diffusion P-waves free of kinetic energy, but as they trigger thermo-mechanical cross-diffusion shear waves  (diffusion S-waves), they can tap into a significant portion of the stored elastic energy around the fault. This process may, for a critical set of reaction-diffusion parameters, ultimately lead to a substantial energy release sufficient to communicate via cross-diffusion with their elastic counterparts \citep{Cartwright}, and an earthquake instability occurs.

Numerical solutions of this stiff problem are difficult, and it may be useful to consider weak-formulations by a reduction procedure through adiabatic elimination of the fast cross-diffusion process into an effective cross-diffusion formulation by time integration to the slow self-diffusion time scale \citep{Biktashev2016}.  Specifically, the upscaled solutions can be regarded as long-wavelength (infinite/quasistatic time-scale limits) of the travelling cross-diffusion wave solutions. The dominant wave number is described by a critical ratio of effective self-diffusion coefficients. The relationship between the long wavelength soliton cnoidal solution of the  Korteweg-de Vries equation \citep{JCSMD2, Veveakis2015cnoidal} and the quasi-soliton solution with oscillating tails can be derived by the Jacobi elliptic functions method of solution. If the modulus of the Jacoby elliptic function asymptotically approaches to zero the solutions shown in Fig.~\ref{fig:Step3} can be obtained \citep{Wang}.

These quasistatic spatially inhomogeneous solutions are also known as Turing patterns \citep{Turing}. The upscaled equations do not necessarily honour the co-dependence between the diffusion and reaction rate constants of the cross-diffusion process \citep{Hu2019b, Vanag}. Turing patterns appear because around the bifurcation point in the reaction-self-diffusion system, linearly unstable eigenfunctions exist that are growing exponentially with time. In order to regularize the problem, finely tuned non-linear terms need to be introduced in the reaction-diffusion equation or, more conveniently, cross-diffusion-like terms need to be added. We postulate here that THMC-Turing  patterns are indeed the multiscale patterns observed in nature \citep{Sethna}. They offer themselves as an ideal tool for inversion of the effective self-diffusion THMC coefficients and their implicit reaction rates. Interpreting geological structures in terms of these process parameters will allow identification of principal processes underpinning the earthquake mechanism. 

Theoretical considerations on the nature of cross-diffusion waves may help in the design and analysis of laboratory experiments. Quasi-solitons are similar to real solitons in that they can penetrate through each other and reflect from boundaries. Differences are that the amplitudes of the true solitons do not change after impact while the dynamics of quasi-solitons on impact is often naturally seen as a temporary diminution of the amplitude with subsequent gradual recovery \citep{Tsyganov2014}. Another important difference is  that the amplitude and speed of a true soliton depend on initial conditions, while for the quasi-soliton they depend on the material parameters. This property  offers new avenues for earthquake physics. We are currently investigating these wave phenomena in controlled laboratory experiments and attempt to use the unique relationship of amplitude and wave speed dependence on material properties as a diagnostic tool for data assimilation. 

\section{Acknowledgments} 
This work was supported by the Australian Research Council (ARC DP170104550,  DP170104557) and the strategic SPF01 fund of UNSW, Sydney.

\section{References}

\bibliography{Uncertainty.bib}
\end{document}